\documentclass[reprint, amsmath,amssymb, pre]{revtex4-1}

\usepackage{graphicx}% Include figure files
\usepackage{dcolumn}% Align table columns on decimal point
\usepackage{bm}% bold math
\usepackage{color}
\usepackage{cases}
\usepackage{empheq}
\usepackage{subfigure}

\newcommand{\be}{\begin{equation}}
\newcommand{\ee}{\end{equation}}
\newcommand{\av}[1]{\langle #1 \rangle}
\newcommand{\bea}{\begin{eqnarray}}
\newcommand{\eea}{\end{eqnarray}}

\begin{document}

\title{Epidemic spreading in modular time-varying networks}

\author{Matthieu Nadini}
 \affiliation{Department of Mechanical and Aerospace Engineering, New York University Tandon School of Engineering, Brooklyn NY 11201, USA\\
 Dipartimento di Elettronica e Telecomunicazioni, Politecnico di Torino, Corso Duca degli Abruzzi 24, 10129 Torino, Italy}%Lines break automatically or can be forced with \\
 \author{Kaiyuan Sun}
 \affiliation{Laboratory for the Modeling of Biological and Socio-technical Systems, Northeastern University, Boston, USA}
 \author{Enrico Ubaldi}
 \affiliation{Institute for Scientific Interchange, ISI Foundation, Turin, Italy}
 \author{Michele Starnini}
  \affiliation{Departament de F\'{\i}sica Fondamental, Universitat de Barcelona, Mart\'{\i} i Franqu\`es 1, 08028 Barcelona, Spain\\
Universitat de Barcelona Institute of Complex Systems (UBICS), Universitat de Barcelona, Barcelona, Spain}
  \author{Alessandro Rizzo}
  \affiliation{Dipartimento di Elettronica e Telecomunicazioni, Politecnico di Torino, Corso Duca degli Abruzzi 24, 10129 Torino, Italy}
\author{Nicola Perra}%
 \email{n.perra@greenwich.ac.uk}
\affiliation{Centre for Business Network Analysis, Greenwich University, London, UK}

%\affil[+]{these authors contributed equally to this work}

%\keywords{Keyword1, Keyword2, Keyword3}

\date{\today}
\begin{abstract}
We investigate the effects of modular and temporal connectivity patterns on epidemic spreading. To this end, we introduce and analytically characterise a model of time-varying networks with tunable modularity. Within this framework, we study the epidemic size of Susceptible-Infected-Recovered, SIR, models and the epidemic threshold of Susceptible-Infected-Susceptible, SIS, models. Interestingly, we find that while the presence of tightly connected clusters inhibit SIR processes, it speeds up SIS diseases. In this case, we observe that heterogeneous temporal connectivity patterns and modular structures induce a reduction of the threshold with respect to time-varying networks without communities. We confirm the theoretical results by means of extensive numerical simulations both on synthetic graphs as well as on a real modular and temporal network.
\end{abstract}

\maketitle
Network thinking has become a prominent and convenient paradigm to unveil the properties of complex systems~\cite{barabasi12-1,butts09-1}. In general, real networks are i) characterized by heterogeneous statistical distributions; ii) organized in modules/communities; and iii) subject to non trivial temporal dynamics~\cite{newman10-1,gcalda,BBV,santo10-1,holme11-1,holme2015modern}. It has long been acknowledged that such attributes have critical effects on dynamical processes evolving on systems' fabric~\cite{BBV}. In particular, the heterogeneity in the connectivity patterns makes networks extremely fragile to the spreading of infectious diseases and malicious attacks~\cite{alex12-1,havlin-book}. Moreover, the presence of communities might slow down the propagation of a disease or facilitate the spreading of social norms~\cite{Onnela:2007,Karsai:2011,centola10-1,centola2015spontaneous}, while temporal changes in networks' structures might inhibit or facilitate spreading processes evolving at comparable time-scales~\cite{Frasca06,Rocha:2010,Isella:2011,Miritello:2011,perra12-1,karsai13-1,scholtes13-1,lambiotte2014effect,PhysRevE.90.042813,RizzoPRE2014,sun2015contrasting,Rizzo2016innovation,Rizzo2016Ebola,zino2016continuous}. Even from this partial list, an extremely interesting and rich phenomenology emerges, often subject to heated debates. \\
The effects introduced by communities and time-varying connectivity patterns on dynamical processes have been mostly scrutinized separately. However, as few recent works pointed out, the two attributes are deeply connected and their interplay introduces non-trivial effects~\cite{liu_social_2017,artime2017dynamics}. The presence of groups, think for example the interactions network of students in a school, introduces specific dynamics that deeply affect spreading processes~\cite{juliette11-1}. 

\section*{Results}
Here, we study the interplay between modularity (i.e., the presence of communities in the network) and time-varying connectivity patterns. To this extent, we introduce a model of time-varying networks with tunable modularity, able to capture several features of real temporal graphs. We derive an analytical characterization of the model, and we study the behaviour of the Susceptible-Infected-Recovered (SIR) and the Susceptible-Infected-Susceptible (SIS) epidemic processes unfolding on its fabrics~\cite{keeling08-1}. Remarkably, while the presence of tightly connected clusters inhibits SIR processes, it favours the spreading of SIS-like diseases, as the interplay between time-varying and modular properties lower the epidemic threshold in the latter case. Interestingly, similar results have been recently obtained in models of time-varying networks characterised by correlated topological features induced by reinforcement of specific ties~\cite{sun2015contrasting}. We confirm the theoretical picture emerging from synthetic networks by means of extensive simulations on a real word dataset of scientific collaborations within the American Physical Society (APS). Our results contribute to characterize the mechanisms, and their interplay, behind the complex, and often contradictory, behaviour of dynamical processes unfolding on real networks.

\begin{figure}
\centering
\includegraphics[width=7cm]{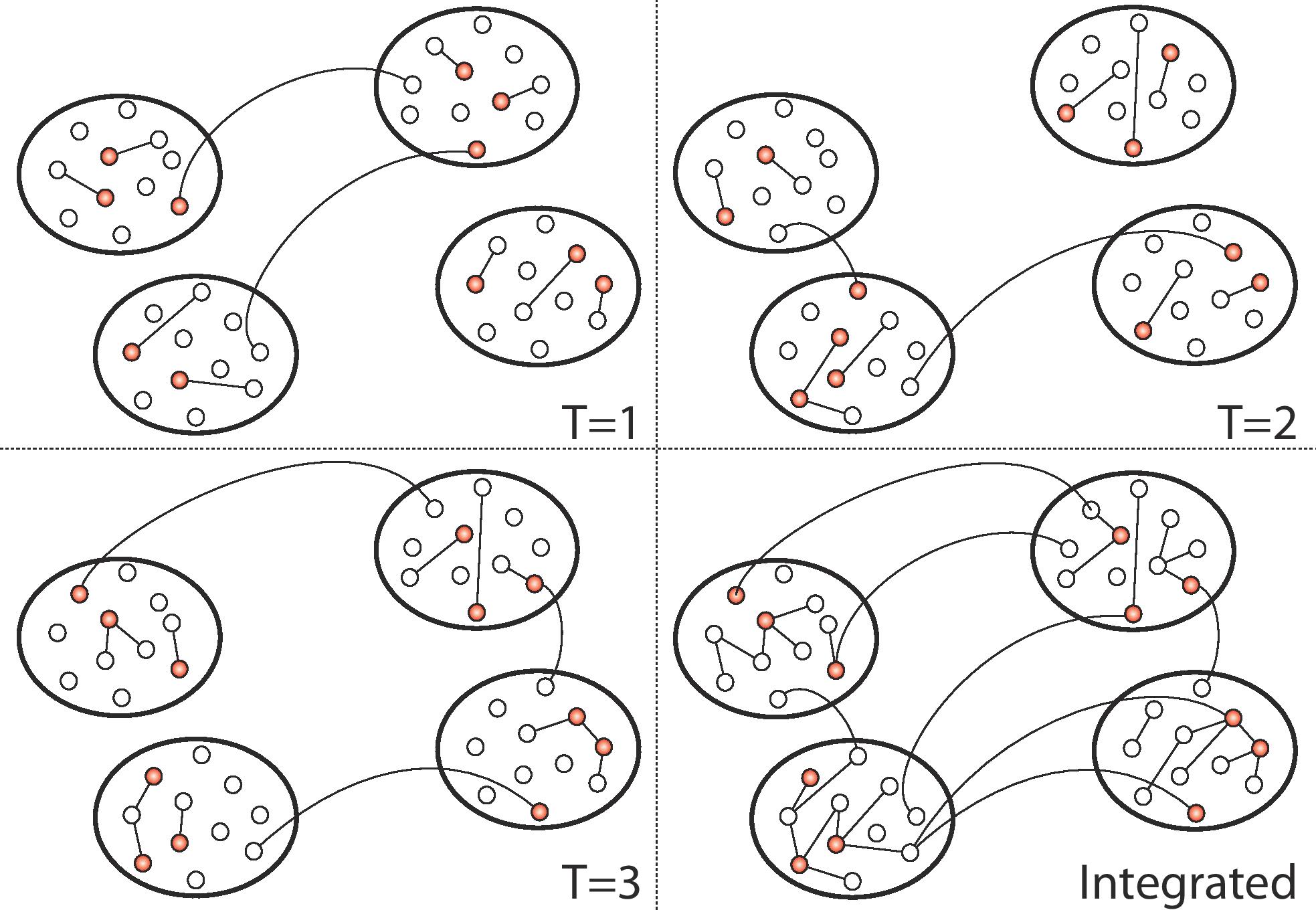}
\caption{Schematic representation of the model. In red, we show active nodes. Straight lines and arcs describe links connecting nodes in the same or in different communities respectively. In the bottom right panel we show the integrated network obtained as the union of $G_1,G_2,G_3$.}
\label{fig:Fig1}
\end{figure}

\paragraph{Modular activity driven networks.}
The system under investigation is composed by $N$ nodes, each characterized by an activity rate $a_i$. This quantity describes the propensity of each node $i$ to engage a social interaction with others. To capture empirical observations performed in a wide set of systems ranging from R\&D to online interactions networks~\cite{karsai13-1,mario14-1,ribeiro12-2,alessandretti2017random}, we consider activity rates extracted from a continuous distribution $F(a)=Ba^{-\nu}$ where $a \in [\epsilon, 1]$ and $\epsilon = 10^{-3}$ to avoid divergence in the distribution. Furthermore, each node is assigned to only one group/community. To consider empirical evidences, the size of each community is extracted from a heavy-tailed distribution, i.e. $P(s)=Cs^{-\omega}$ with $s \in [s_{min},\sqrt{N}]$~\cite{santo10-1,lancichinetti08-1}. Therefore, we do not limit ourselves in studying a fixed number of modules\cite{liu_social_2017}, whilst their number is driven from the model's parameters.\\
Given these settings, a generative network model is defined by the following steps (see Fig.~\ref{fig:Fig1}).
\begin{itemize}
\item At each time $t$, the network, $G_t$, starts with $N$ disconnected nodes.
\item With probability $a_i\Delta t$ each vertex $i$ is active and willing to create $m$ connections.
\item With probability $\mu$ each link is  generated within the node's community, and with probability $1-\mu$ with nodes in any other groups. In both cases nodes are selected randomly.
\item At the next time step $t+\Delta t$ all the edges in $G_t$ are deleted.
\end{itemize}
All the interactions have a constant duration $\Delta t$. In the model, neither self-loops nor multiple edges are allowed. In the following, without loss of generality, we fix $\Delta t=m=1$.

At each time step, the model generates a random, structureless, network in which few nodes are active. The modular features of the network emerge integrating connections in time. Such time-integrated properties, at different time regimes, can be computed analytically. In the following, we will report the results only for the evolution of the average number of connections of each node $\av{k_i(t)}$ (average degree) and the overall degree distribution $\rho(k)$ (for the complete set of results see the Supplementary Information).

To solve the average degree's dynamics, let us introduce the effective activity $\tilde{a}_i = a_i + \av{a}$ and the mixing parameter $\mu^{\prime} = 1-\mu$. We refer to the degree of node $i$ at time $t$ as $k(a_i,s,t)$, where $s$ is the node's community size. By defining an activity class as the group of nodes featuring similar activity values $a$, we set the average in-community degree $\av{k_c(a,s,t)}$ to be the average number of connections that nodes belonging to the activity class $a$ and falling in communities of size $s$ have toward nodes of their same community. The latter grows as
\begin{equation}
\left\langle k_c (a,s,t) \right\rangle = (s - 1) \left[1 - \exp\left(- \frac{t}{\tau(a, s)} \right)\right],
\label{k_c}
\end{equation}
where $\tau(a,s)$ is the characteristic time that it takes for the degree $k_c(a,s,t)$ of nodes of activity $a$ belonging to a community of size $s$ to be $k_c(a,s,t)\sim (s-1)$, being $s-1$ the maximum value of the in-community degree (see the Supplementary Information for the evaluation of $\tau(a,s)$).

Similarly, we can define the average out-community degree $\left\langle k_o(a,t) \right\rangle$ as the number of connections that nodes of activity class $a$ have outside of their communities at time $t$. We expect this quantity to be independent on the nodes' communities size $s$ so that, for large networks we can write:
\begin{equation}
\left\langle  k_o (a,t)\right\rangle= \mu^{\prime} \tilde{a}t
\label{k_o_eq}
\end{equation}

The average total degree $\av{k(a,s,t)}$ can be computed as the simple sum between the two previous equations, obtaining
\begin{widetext}
\begin{subnumcases}{\av{k(a,s,t)} = \av{k_c(a,s,t)} + \av{k_o(a,t)}\simeq}
   \tilde{a}t & $t\ll \tau(a, s)$ \label{k_T short time} \\
   \mu^{\prime} \tilde{a}t + (s-1) & $t\sim \tau(a, s)$ \label{k_T_similar_times} \\
   \mu^{\prime} \tilde{a}t & $t\gg \tau(a, s)$ \label{k_T_big_times}
\end{subnumcases}
\end{widetext}
Three regimes are readily identified: an initial growth in which both the in-community and the out-community degrees are growing linearly in time, followed by the slowing down of the in-community degree, which saturates to $s - 1$, and then a further linear regime driven only by the out-community degree growth. Fig. \ref{k_tot_av} shows that the numerical simulations perfectly match with the theoretical formulas (see the Supplementary Information for details).\\
\begin{figure}
\centering
\includegraphics[width=8cm]{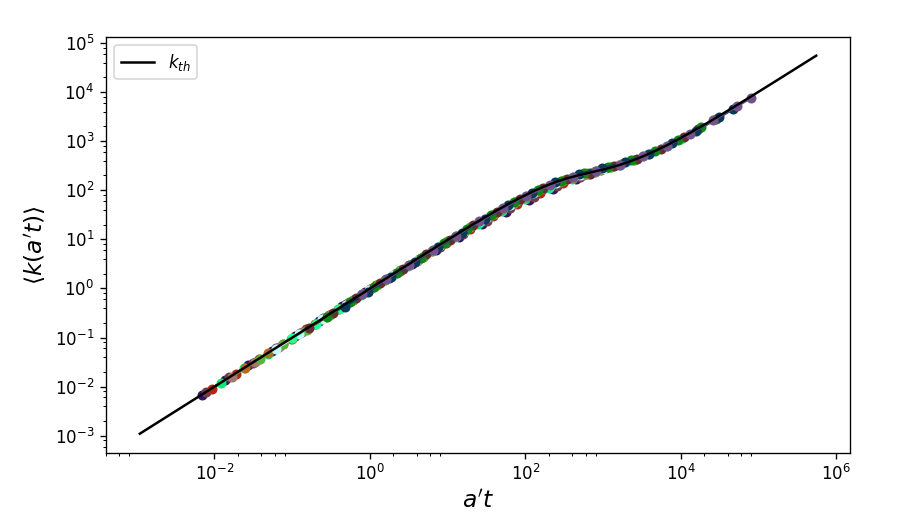}
	\caption{
    	Time evolution of the average total degree, $\av{k(a,s,t)}$, for different activity classes and compared with the theoretical function of Eqs. \ref{k_T short time}, \ref{k_T_similar_times} and \ref{k_T_big_times}, evaluated considering a community size equal to the average (i.e. $s=\av{s}$). The rescaled time is $t \rightarrow \tilde{a}t$ and $\av{k(\tilde{a}t)}$ is plotted. Parameters used are: $N = 10^5$, $\omega = 2.1$, $\nu = 2.1$, $m=1$, $s_{min}=10$, $\mu = 0.9$ and $T=10^5$ evolution steps. Each point is an average of $10^2$ simulations.
    }
	\label{k_tot_av}
\end{figure} 

\begin{figure}
\centering
\includegraphics[width=6cm]{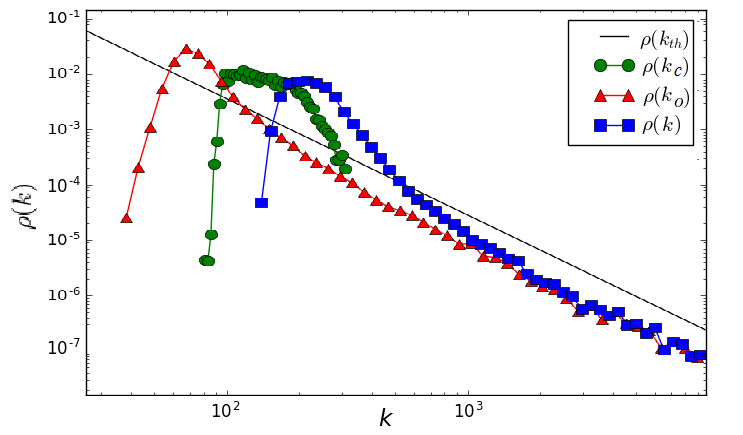}
\caption{Plot of the three degree distributions and the theoretical prediction, given in Eq. \ref{degree_distributions}. Parameters used are: $N = 10^5$, $\omega = 2.1$, $\nu = 2.1$, $m=1$, $s_{min}=10$, $\mu = 0.9$ and $T=10^5$ evolution steps.}
\label{degree_distribution_plot}
\end{figure}

Noticeably, the long time evolution of the node degree is linear in time and proportional to its activity class $a$, so that we find the asymptotic degree distribution of the system to feature the same functional form of $F(a)\propto a^{-\nu}$:
\begin{equation}
F(a)da \xrightarrow[]{k(a,t) \propto a\cdot t} \rho(k) dk \propto k^{-\nu} dk.
\label{degree_distributions}
\end{equation}
In Fig. \ref{degree_distribution_plot}, we integrate the network for $T=10^5$ and we plot the three degree distributions. As expected, the out-community $\rho(k_o)$ and the total $\rho(k)$ degree distributions falls as power laws with exponent $-\nu$. On the other hand, the in-community degree $\rho(k_c)$ saturate to the community size distribution $P(s)$, as all the nodes reach their maximum in-community degree value $(s-1)$, being that the modules' size is far smaller than the network size ($s_{\text{max}}=\sqrt{N} \ll N$). On the contrary, the out-community degree takes longer times to saturate to its maximum value $N-s \gg s$.\\
It is worth stressing that the results presented in this section apply to the networks obtained integrating links over time. A process unfolding on such networks, in general, will be affected by the time-aggregated features of the graph. The extent to which this is true, is function of the interplay between the time-scale describing its evolution, $\tau_{P}$, and the various $\tau(a, s)$. In the limit $\tau_{p}\ll \tau(a, s)$ the process would effectively evolve on the instantaneous, annealed  networks that are characterized by a small average degree and modularity. In the opposite limit instead, the process would effectively unfold on static networks  obtained integrating links over longer time characterized by high average degree and low modularity. Indeed, the average degree is this regime will be dominated by out-community links that make the connections between different communities increasingly stronger, thus increasingly destroying the identity of communities. In the limit  $\tau_{p}\sim \tau(a, s)$ the process would effectively evolve on maximally modular networks (for a given set of parameters). Arguably, this is the most interesting regime that we will consider in the following.

\paragraph{Epidemic processes on modular activity driven networks.}
Let us turn our attention on the dynamical properties of SIR and SIS processes (see the Methods section for a detailed definition of the two) unfolding on the proposed model. Although similar, the two processes are intrinsically different~\cite{ferreira12-1,castellano10-1,goltsev12-1,k14-1}. Indeed, SIR processes are always characterized by the so called disease-free equilibrium, provided $d_t N=0$.
The illness eventually disappear, i.e., $I=0$ for $t\rightarrow \infty$. SIS models instead allow the existence of an endemic state where a finite and constant fraction of infected individuals permanently colonize the population, i.e., $I>0$ for $t\rightarrow \infty$.\\
We focus on a central concept of contagion phenomena: the epidemic threshold. This quantity defines the conditions necessary for the  spreading of the illness. In annealed networks the threshold is determined by the moments of the degree distribution $P(k)$, that specify the probability of finding a node with $k$ distinct neighbours~\cite{alex12-1}. In static graphs the expression is given by the principle eigenvalue of the adjacency matrix $\mathbf{A}$, defined as $A_{ij}=1$, if $i$ and $j$ are connected, and  $A_{ij}=0$ otherwise~\cite{wang03,castellano10-1,durrett10-1}. In time-varying networks instead, the threshold is determined by the interplay between the time-scales of the contagion and network evolution processes~\cite{prakash10-1, valdano2015analytical,starnini13-1,lee12-1,takaguchi12-1,tang11-1,Masuda13-1,perra12-1,liu13-2,RizzoPRE2014,zino2016continuous,pozzana2017epidemic}. In the case of SIR models, we also consider another important quantity: the epidemic size $R_\infty$ which is defined as the final ratio of recovered nodes. This describes the fraction of nodes affected by the disease. 

\begin{figure}
\centering
\includegraphics[width=8cm]{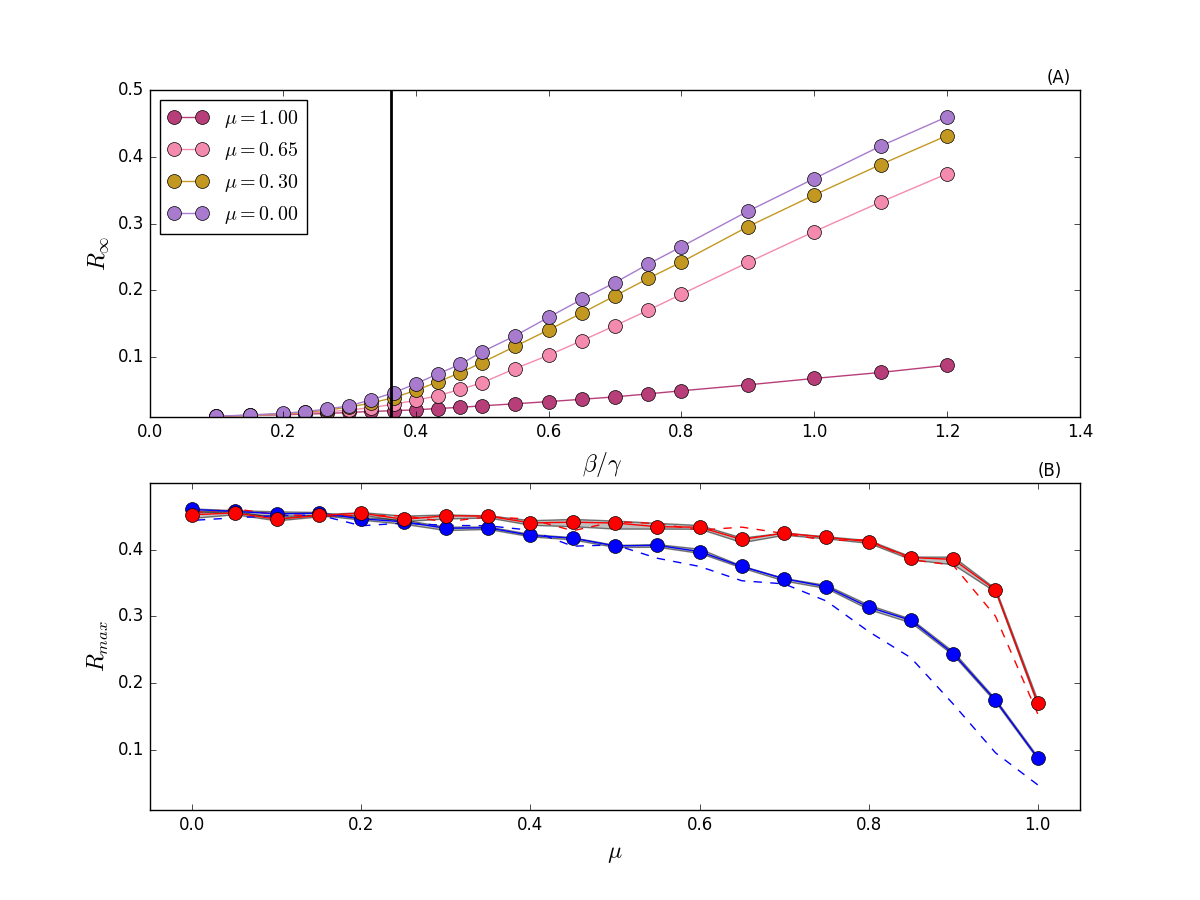}
\caption{Panel A) $R_{\infty}$ as a function of $\beta/\gamma$, for selected values of $\mu$ and $s_{\text{min}}=10$. Vertical black line represents the theoretical value of the epidemic threshold for $\mu = 0$ as derived in Refs.~\cite{perra12-1,starnini14-1}. Panel B) $R_{\text{max}}$, i.e. the max value of $R_{\infty}$, as a function of $\mu$. In red curves we set $s_{\text{min}}=100$, in  blue curves $s_{\text{min}}=10$. In solid curves, we draw community sizes directly from the community size distribution $P(s)$. In dashed curves, we fix the community sizes as equal to the average value of $P(s)$ for all communities.  The 95\% confidence interval is in gray. Each point is an average of $10^2$ independent simulations.\\}
\label{SIR}
\end{figure}

\begin{figure}
\centering
\includegraphics[width=8cm]{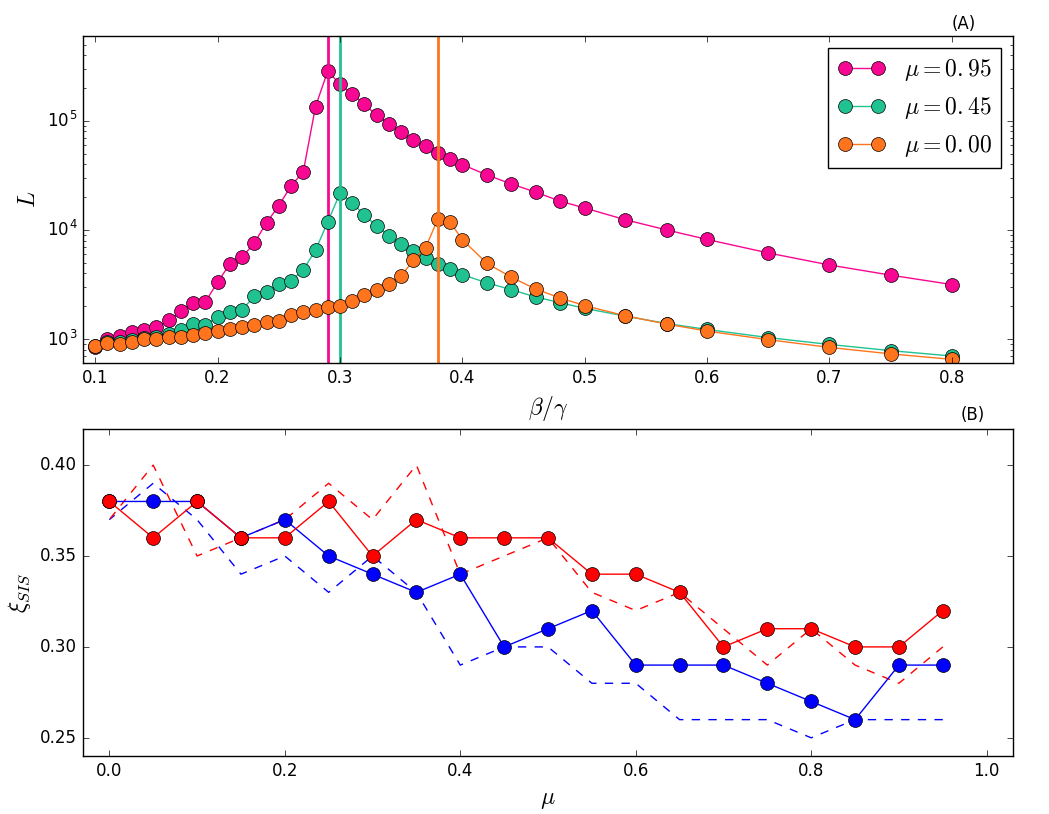}
\caption{Panel A) Lifetime of the disease $L$ as a function of $\beta/\gamma$, for selected values of $\mu$ and when $s_{\text{min}} = 10$. Vertical lines are the epidemic threshold. Panel B) Ratio $\xi_{\text{SIS}}=\beta/\gamma$ in correspondence of $L_{\text{max}}$, as a function of $\mu$. In red curves we set $s_{\text{min}}=100$, blue curves $s_{\text{min}}=10$. Each point is an average of $10^2$ independent simulations. Note that we avoid to simulate $\mu = 1$ because the criterion we follow for the estimation of the threshold does not hold for a network with many connected components.}
\label{SIS}
\end{figure}

To develop a deeper understanding, let us derive the mean-field level dynamical equations describing the contagion process in modular activity driven networks. We define the activity block variables $S_{a,s}$, $I_{a,s}$, and $R_{a,s}$  as the number of susceptible, infected and recovered individuals, respectively, in the class of activity $a$ and community of size $s$ at time $t$ (to enhance readability, we omit to notate the dependence on time). This allows us to write the mean-field evolution of the number of infected individuals, for a SIR process, in each group of nodes with activity $a$ as:
\begin{eqnarray}
\label{eq1}
d_tI_{a,s} &=&-\gamma I_{a,s}+\lambda S_{a,s}\left [ \mu a \frac{I_{s}}{s}+(1-\mu)a\frac{I}{N}\right]  \nonumber \\
&+&\lambda \sum_{a'}a'\left [ \mu I_{a',s}\frac{S_{a,s}}{s} +(1-\mu) I_{a',s}\frac{S_{a,s}}{N}  \right],
\end{eqnarray}
where $I_s$ and $I$ are the number of infected in communities of size $s$ and in the whole network, respectively.  The first term in the r.h.s accounts for the recovery of infected individuals. The other four terms account for the probability that a Susceptible node in a community of size $s$ connects to an Infected node inside (first) or outside (second) its community acquiring the infection, and for the probability that an Infected node of class $a'$ connects to a Susceptible node inside (third) or outside (forth) a community of size $s$, contracting the disease. For simplicity, we consider that $N-s\sim N$ and, at least initially, $I-I_{s}\sim I$. Summing over all the activities and community sizes, and considering only the first order terms in $a$, $I_{a,s}$, $R_{a,s}$ and their products, we obtain 
\begin{eqnarray}
\label{eq2}
d_t I&=&-\gamma I + \lambda \av{a}I +\lambda \Theta +\lambda \mu \sum_s (\av{a}_s-\av{a})I_s, \\
d_t \Theta &=&-\gamma \Theta +\lambda  \av{a^2}I +\lambda\av{a}\Theta + \nonumber \\
&+& \lambda \mu \sum_s \left [ (\av{a^2}_s-\av{a^2})I_s + (\av{a}_s-\av{a})\Theta_s \right],
\end{eqnarray}
where we defined $\Theta =\sum_a{a}I_a$,  and $\Theta_s=\sum_{a}aI_{a,s}$. The term $\av{a^x}_s=\sum_aN_{a,s} a^x/s$ describes the moments of the activity distribution in any community of size $s$. The second, auxiliary, equation is obtained from the first  by multiplying both sides by $a$ and summing over all $s$ and $a$. The epidemic threshold, in principle, can be derived evaluating the principle eigenvalue of the Jacobian matrix of the system of differential equations in $I$ and $\Theta$~\cite{perra12-1,liu13-1,liu13-2,RizzoPRE2014,zino2016continuous,pozzana2017epidemic}. In general, a closed expression for the threshold does not exist. However, we can point out some interesting observations. First of all, the terms associated to $R_{a,s}$ vanish, implying that, at the first order, the thresholds of both SIR and SIS are equal~\cite{liu13-2}. Furthermore, the terms in $\mu$ weigh a comparison between the moments of the activity distribution in the network with the corresponding quantities evaluated inside each community. In realistic cases, where $s \ll N$, fluctuations act differentiating between these values. Instead, if they are negligible, due for example to very large community sizes or to narrow distribution of activity, the equations become equivalent to the case $\mu = 0$. In the limit $\mu \rightarrow 0$ the network has no modular structure. The threshold, for both SIR and SIS, becomes $\beta/\gamma \ge 2/(1+\sqrt{\chi})$ as derived with different approaches in Refs.~\cite{perra12-1,starnini14-1,RizzoPRE2014,zino2016continuous}. We defined $\chi=\av{a^2}/\av{a}^2$, where the moments are evaluated over the whole network. As expected, the spreading condition is determined by the interplay between the time-scale of the contagion process and  the time-scales of the network. In the opposite limit $\mu \rightarrow 1$ networks are extremely modular. Fluctuations become important and the symmetry between SIR and SIS breaks. In order to understand this limit, let us consider first a SIR process started from a single infected node in a community of size $s$. The large majority of connections are towards vertices in the same group. As soon as some infected node recover, the probability of links connecting $I$ and $R$ nodes increases. Such connections hamper the spreading of the disease. From these simple observations we can expect that SIR processes are inhibited by highly modular connectivity patterns. On the other hand, in case of SIS processes, the repetition of contacts does not lead to such ''pair annihilation": contacts between infected nodes do not help the spreading of the disease, but they are only temporary (eventually, all infected nodes become susceptible again). Thus, we  expect that modularity plays a different role in SIS dynamics. \\
In order to numerically characterize SIR models, we study the epidemic size, $R_\infty$, as a function of $\beta/\gamma$. Indeed, this quantity acts as order parameter of a second-order phase transition~\cite{alex12-1}. For SIS processes instead, the order parameter is the final fraction of infected individuals, $I_\infty$~\cite{alex12-1}. The numerical estimation of this quantity is challenging, since it requires the precise determination of endemic states. For these reasons, we follow Ref.~\cite{boguna13-1}, measuring the life time of the disease, $L$, that acts as the susceptibility in phase transitions~\cite{aharony2003introduction}. This quantity is defined as the average time the disease takes to either die out or reach a macroscopic fraction, $Y$,  of the populations. Without loss of generality, we start our simulations by setting $1\%$ of randomly selected nodes as initial infected seed. Other parameters are set as: $\gamma=0.01$, $m = 3$, $\nu = 2.1$, $\omega = 2.1$, $N=10^5$ and  $Y=0.5$ (see SI for similar plots obtained fixing $\omega=1.5$).\\
Results obtained from SIR models are represented in Fig. \ref{SIR}A-B, whilst results from SIS models are visible in Fig. \ref{SIS}A-B. In Fig. \ref{SIR}B and \ref{SIS}B we study different community structure, either by considering a constant community size (dashed curves) or by drawing  community sizes directly from the community size distribution $P(s)$ (solid curves). In general, red curves represents a network with bigger communities than the one represented with blue curves.\\
For SIR models, Fig. \ref{SIR}A tells us that, as expected, the higher $\beta/\gamma$ the higher the epidemic size. Interestingly, we observe a weak dependence of the threshold on $\mu$. Moreover, the higher the fraction of links created between pair of nodes sharing the same community (i.e. the higher $\mu$), the lower the epidemic size. This second observation is confirmed studying different community structures, as done in Fig. \ref{SIR}B, in which we plot the maximum epidemic size (corresponding to the largest value of $\beta/\mu$ in our settings), $R_{\text{max}}$, as a function of $\mu$. In the limit $\mu \rightarrow 0$, we observe that the disease impact is the same: the networks behave as if no community structure was present. Instead, when $\mu \rightarrow 1$, the modular structure influences the spread of the disease. As mentioned before, repeating contacts within communities  significantly narrows the chances of having new infected individuals. Indeed, in SIR models, once a node recovers, it cannot be infected again. Repeating contacts with nodes already recovered does not favor the spread of the disease. Overall, the main observations are four. (i) Increasing the modularity reduces the epidemic size. (ii) A network with, on average, larger modules is likely to yield a higher epidemic size. (iii) The larger the modules the weaker the dependence on $\mu$ of the epidemic size. (iv) In case of small modules, the distribution of communities size seems to influence the spreading of the disease. In particular, a network organized in small groups of constant sizes leads to smaller epidemic size respect to a network in which the average community size is the same, but individual sizes are extracted from a power-law distribution.\\
For SIS models, the lower $\mu$, the lower the life time $L$ (see Fig. \ref{SIS}). Inter-community links speed up the disease spreading and an endemic state, i.e. $Y = 0.5$, is reached faster. Moreover, the higher $\mu$, the lower the epidemic threshold. This last observation, which implies that increasing values of modularity favor the survival of the disease, is confirmed in Fig. \ref{SIS}B where we also test the effects of  different community structures. In the limit $\mu \rightarrow 0$, there is no community structure and the curves converge to the same epidemic threshold. On the contrary, when $\mu \rightarrow 1$, the community structure becomes increasingly important and influences the spreading. Qualitatively, higher levels of modularity diminish the epidemic threshold. This is due to the repetition of the same contacts within a community which becomes increasingly more likely. Indeed, in SIS models, reinfection is allowed and nodes can become infected many times: communities act as a reservoir for the disease and favor the contagion process pushing the epidemic threshold to smaller values.  Besides this last point, there are two main observations. (i) A network with larger modules is likely to have an higher epidemic threshold. (ii) In case of communities with smaller average sizes and high values of modularity, having communities sizes extracted from a power-law seems to slightly increase the threshold. Thus, the disease is able to spread more easily in modular networks with communities of similar or equal sizes. With the exception of one data point, this is observed for $\mu>0.5$ (see the dashed blue line in Fig. \ref{SIS}-B).  \\
To summarize: in the limit $\mu \rightarrow 0$ the community structure becomes irrelevant and the disease spreads as if no modules are present. On the contrary, when $\mu \rightarrow 1$, repetition of same contacts within communities favors the contagion process in SIS models and slows it down in SIR models.

\begin{figure}
\centering
\includegraphics[width=8cm]{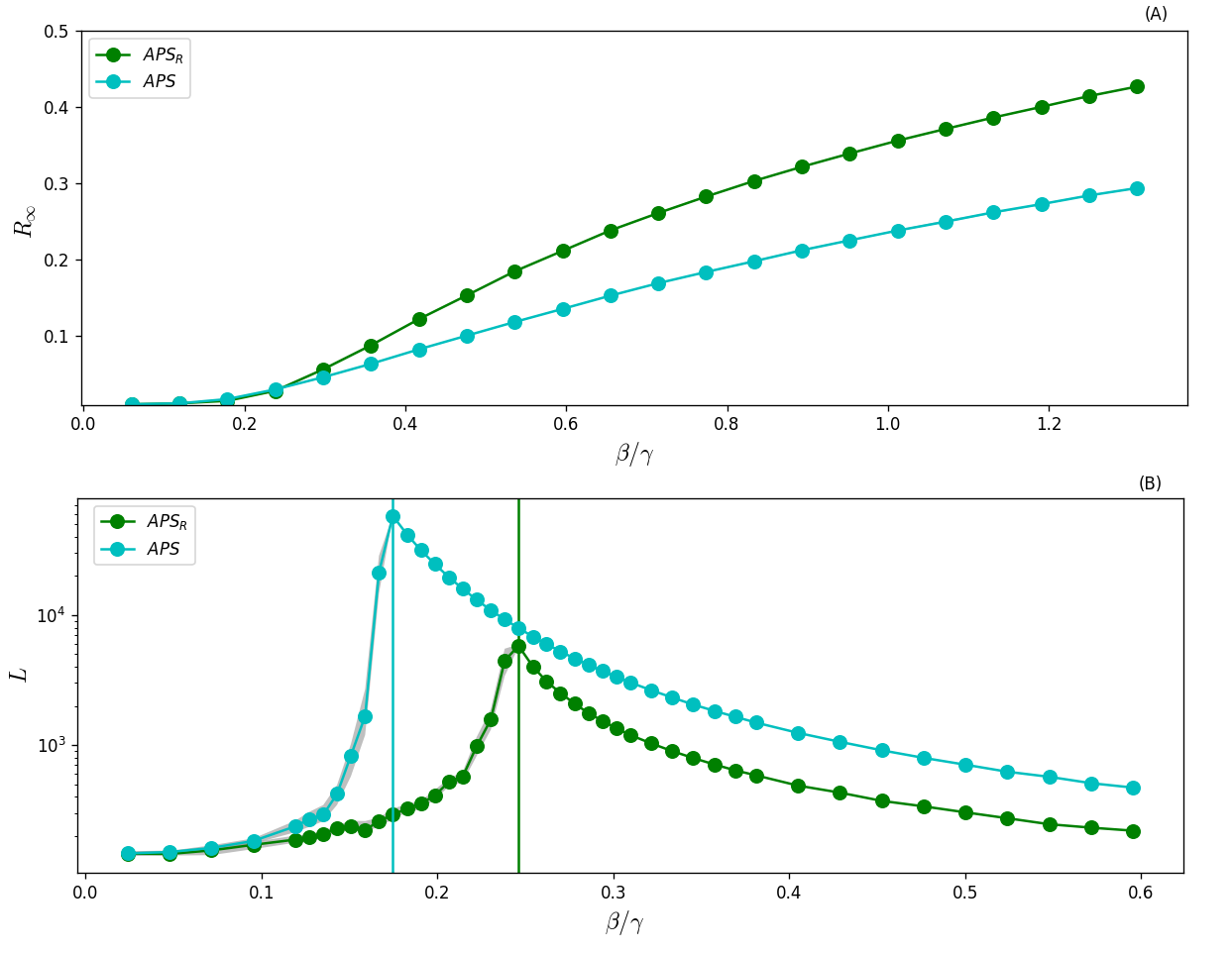}
\caption{Panel A) $R_\infty$ as a function of $\beta/\gamma$ for SIR processes diffusing on APS (cyan circles) and on the randomized APS dataset (green circles). Panel B) $L$ as a function of $\beta/\gamma$ for a SIS models evolving on the same two networks. Each point is the average of $10^2$ independent simulations started from $1\%$ of random seeds. We fix $\gamma=0.05$.}
\label{fig:Fig_real}
\end{figure}

\paragraph*{Real networks.}
Although the modelling framework presented captures realistic activity and community size distributions of real networks, it neglects other important features such as burstiness \cite{ubaldi2017burstiness,
goh2008burstiness, moinet2015burstiness,Lambiotte2013,karsai2012universal}, and more complex temporal/structural correlations \cite{tiago2017,laurent2015calls,pfitzner2013betweenness,vestergaard2014memory,ubaldi2016asymptotic}. It is then crucial grounding the picture emerging from synthetic models with a real world system. To this extent, we consider a temporal and modular  network about scientific collaborations in the American Physical Society (APS). We study $96940$ scholars connected by $692667$ links (see the Supplementary Information for more details)~\cite{SARA}. We focus on ten years of data (January 1997 - December 2006) coarse-grained at a time resolution of one month. To single out the effects introduced by communities on contagion processes, we consider also a randomized version of the dataset. Here, the interactions at each time are shuffled, modules are destroyed, but the sequence of activation times for each node and the degree distribution at each time step are preserved~\cite{starnini_rw_temp_nets}. In order to make sure that the randomization process removes topological structures, we integrate the two networks over all time steps and we use OSLOM\cite{lancichinetti_finding_2011} to find the communities. The modularity\cite{newman_finding_2004} of the real APS network is $Q=0.6685$, and of its randomized counterpart $Q =0.0937$. As expected, the degree preserving randomization reduces the modularity significantly. Using these two networks, we study the dynamical properties of SIR and SIS processes unfolding on their structure. In Fig.~\ref{fig:Fig_real}A-B we present the results. Interestingly, the modular properties of the real network do not influence the threshold of SIR models. Nevertheless, the presence of communities reduces the impact of the disease, i.e. lowers the epidemic size. In the case of SIS processes instead, communities have a larger effect shifting the threshold to smaller values. These results qualitatively confirm what observed in synthetic systems. 

\section*{Discussion}
We have presented a model of temporal networks with tunable modularity and heterogeneous activity distributions. We have provided an analytical characterization of time-aggregated properties of such networks. Within this framework, we have studied the interplay between modularity and temporal dynamics. In synthetic networks, we have found that modularity reduces the epidemic size in SIR models, slowing down the spreading process. On the other hand, in SIS models, modularity reduces the epidemic threshold making the system more prone to disease spreading. Indeed, repetition of the same contacts between nodes sharing the same community acts as a reservoir for SIS-like diseases and allows the pathogen to reach an endemic state more easily. Modular activity-driven networks do not capture all crucial aspects of real time-varying networks, as the appearance of new nodes, disappearance of old ones, bursty behaviours. The introduction of these features is left for future works. However, we studied SIR and SIS spreading in a real modular and temporal network confirming the picture emerging from synthetic graphs.\\
In conclusion, the results here presented show that the interplay between modularity and temporal dynamics can have opposite effects on different classes of
spreading processes. Our findings contribute towards the efforts aimed at characterising how spreading processes are affected by the features of real networks.

\section*{Methods}

\paragraph*{SIS and SIR models.} In both processes nodes are divided in different classes according to their disease status. In SIR models nodes are either Susceptible (S), Infected (I) or Recovered (R). Susceptible nodes describe healthy individuals. Infected nodes contract the disease and are infectious. Recovered nodes are no longer infected and acquire complete immunity to the illness. The model is fully characterized by two transitions: $S+I\xrightarrow{\beta} 2I$ and $I\xrightarrow{\gamma} R$. The first describes the infection propagation. Susceptible nodes in contact with infected individuals become infected with rate $\beta$. This quantity is defined by the average contacts per node $\av{k}$ and by the per contact probability  of transmission $\lambda$, i.e $\beta=\lambda \av{k}$. The second transition describes the recovery process. Infected individuals recover spontaneously and permanently with rate $\gamma$. In SIS models instead we have just Susceptible and Infected nodes. While the contagion process is equivalent to the SIR case, the recovery is different and described by the following transition: $I\xrightarrow{\gamma} S$. Infected nodes spontaneously return in the susceptible compartment with rate $\gamma$.

\bibliography{refs}

\section*{Acknowledgements}
M.S. acknowledges financial support from the James S. McDonnell Foundation.
M.N. thanks the Centre for Business Networks Analysis at the University of Greenwich for support and hospitality during this project. M.N. and A.R. acknowledges financial support from the National Science Foundation under grant No. CMMI-1561134 and the Army
Research Office under grant No. W911NF-15-1-0267, with Drs. A. Garcia and S.C. Stanton as program
managers. A.R. acknowledges financial support from Compagnia di San Paolo, Italy. 

\section*{Author contributions statement}

N.P. conceived the research,  M.D. and K.S. conducted the numerical simulations, E.U. developed the analytical calculations, all authors analysed the results, wrote and reviewed the manuscript. 

\section*{Additional information}

We now present in a more detailed and comprehensive way the definition of the
model, its key properties and the results (both analytical and
numerical) found. Furthermore, we show additional simulations studied in synthetic networks for SIR and SIS models. Finally, an explanation of the APS dataset  is given.

\section{The Model} % (fold)
\label{sec:The Model}
The network is defined by means of the following parameters:
\begin{itemize}
    \item [-] the total number $N$ of nodes in the network;
    \item [-] the activity distribution parameters, i.e. the lower cut-off
        $\epsilon$ and the leading exponent $\nu$, so that $F(a) \propto a^{-\nu}$
        for $a\in [\epsilon, 1]$;
    \item [-] the lower cut-off $s_{\rm min}$, the upper limit $s_{\rm max}$ and
        the exponent $\omega$ describing the community size distribution,
        i.e. $P(s) \propto s^{-\omega}$ for $s \in [s_{\rm min}, s_{\rm max}]$;
    \item [-] $\mu$ is the probability that, once active, a node will connect to a node inside
        the community, so that $\mu^{\prime} = 1 - \mu$ is the probability to fire outside
        the community;
\end{itemize}

We initialize the network extracting $N$ activity values from the activity distribution $F(a)$ and we then group the nodes in communities of size $s$ drawn from a size-distribution $P(s)$. Once we initialized the network we let it evolve following the time-varying activity driven framework. At each time step $t$ we start with $N$ disconnected nodes. Each node gets active with probability $a_i dt$ at each time step $dt$ and fire to a randomly chosen node inside (outside) its own community with probability $\mu$ ($\mu^{\prime}$). At time $t+dt$ we delete all the edges and repeat the above procedure. Each node $i$ will then have a set of neighbors that have been contacted or have contacted the node during the network growth. The size of such a set is the integrated degree $k$ of the node $i$. Of these $k$ neighbors, $k_c$ will be inside the $i$'s community (in-community degree) and $k_o=k-k_c$ will be external to the community (out-community degree).

% section The Model (end)

\section{The Network growth} % (fold)
\label{sec:Network_growth}

In the integrated network, each node $i$ has a set of neighbors that have been contacted or have contacted the node during the network growth. The size of such a set is the integrated degree $k$. Of these $k$ neighbors, $k_c$ are inside the $i$’s community (i.e. the in-community degree) and $k_o = k - k_c$ are external to the community (i.e. the out-community degree). Since the model is memoryless, the in-community degree $k_c$ and the out-community degree $k_o$ are decoupled and can, in fact, be treated separately.\\
Even the activity potential $a_i$ of each node can be ``split'' in two components: the in-community activity $\mu a_i = (1-\mu^{\prime})a_i$ and the complementary out-community $\mu^{\prime} a_i$. Indeed, each node points, on average, a fraction $\mu$ of its own events toward the community, while the remaining $\mu^{\prime}$ are directed outside the community itself. Each node experiences a mean field of activity, $\mu \left\langle a \right\rangle$, coming from the community (provided that the community is large enough) and a supplementary external field $\mu^{\prime} \left\langle a \right\rangle$ coming from the rest of the network.

\subsection{The network time scales} % (fold)
\label{sub:The network time scales}

As a first insight, let us note that the in-degree time dependence can be easily approximated with a probabilistic consideration. Each node $i$ of activity $a_i$, within a community of size $s$, has $s - 1$ available edges. Now, for each
time step of the dynamics, the edge $e_{ij}$ is created with probability $\mu(  a_i + a_j)/(s - 1)$. Then, on average, each edge emanating from $i$ is activated with probability $c(a_i, \mu)/(s - 1) = \mu(a_i + \left\langle a \right\rangle)/(s - 1)$, where $c(a_i, \mu)$ is the number of edges intra-community. The probability $P(a_i, \mu, s, t)$ for an edge pointing to $i$ not to be activated after $t$ time steps then reads:

\begin{equation}
P(a_i, \mu, s, t) = \left( 1 - \frac{c(a_i, \mu)}{s-1} \right)^t.
\label{Edge_probability_supp}
\end{equation}
Since the in-degree $k_c \in [0, s-1]$, we can write:

\begin{equation}
k_c(a_i, \mu, s, t) = (s - 1) (1 - P(a_i, \mu, s, t)).
\label{k_c_supp}
\end{equation}
In the following, we always have the dependency on $a_i$, $\mu$ and $t$, so to simplify the notation we drop most of those parameters: $P(a_i, \mu, s, t) = P^{\prime}(s)$, to avoid confusions with the community size distribution P(s). Also, $k_c(a_i, \mu, s) = k_c(s)$ and $c(a_i, \mu) = c$.  

We note that Eq. \ref{Edge_probability_supp} gives us an estimation of the characteristic time $\tau (s)$ that takes for a node of activity $a_i$ to saturate the in-degree $k_c \rightarrow (s - 1)$. Indeed, we can rewrite Eq. \ref{Edge_probability_supp} as:
\begin{widetext}
\begin{equation}
P^{\prime}(s)= \exp\left[t \ln\left( 1 - \frac{c}{s-1} \right) \right] \Rightarrow \tau(s) = - \left[\ln \left(1-\frac{c}{s-1} \right) \right]^{-1}.
\label{tau_s_supp}
\end{equation}
\end{widetext}
So, as expected, the saturation time (i.e. the typical time for $k_c$ to be of the same order of $s$) increases as the activity $\mu a_i$ decreases and/or the community size $s$ grows.\\
Generalizing the above reasoning, the characteristic time for a community to have the majority of the nodes saturated is obtained by evaluating the probability $P_e(s)$ to create  (on average) an edge $e_{ij}$ in a community of size $s$ in a single evolution step. In other words, it is the number of edges activated in one step divided by the total number of possible edges in the network:
\begin{equation}
P_e(s)=\frac{2 s \mu \left\langle a \right\rangle}{s(s-1)}.
\end{equation}
The probability for one edge not to be created after $t$ time steps is then:
\begin{widetext}
\begin{equation}
\begin{aligned}
\bar{P}_e(s)=\left(1-\frac{2\mu \left\langle a \right\rangle}{s-1} \right)^t & \Rightarrow \bar{P}_e(s) = \exp\left[- \frac{t}{\tau_c(s)} \right] \Rightarrow \tau_c(s) = - \left[\ln \left(1- \frac{2\mu \left\langle a \right\rangle}{s-1} \right) \right]^{-1},
\end{aligned}
\end{equation}
\end{widetext}
where $\tau_c(s)$ represents the typical time by which the majority of the nodes of a community has a degree $k_c \simeq s$.

Note that, in the evaluation of both $\tau(s)$ and $\tau_c(s)$, we did not take into account the difference between edges pointing to a more active node and the ones pointing to a less active one. Nevertheless, this is a simple estimation that, as we will show later, correctly catches the general behaviour of the in-degree $k_c$ for any value of $\mu$, $a_i$ and $s$.
Besides, when computing the key features of the evolving network, we are now able to distinguish the short time range $t \ll \tau_c(s)$ (in which $k_c \ll s$ for any activity value $a_i$) and the long time limit $t \gg \tau_c(s)$ (in which $k_c \sim s$ for any activity value $a_i$). 

% subsection The network time scales (end)

\subsection{The Master Equation and the $P(a, k, t)$} % (fold)
\label{sub:ME}

We can now write down the Master Equation (ME) for the quantities $P_c(s, k_c)$ and $P_o(s, k_o)$, that is, the probability for a node of activity $a_i$ belonging to a community of size $s$ to have degree in (out) degree $k_c$ ($k_o$) at time $t$. In general, $a_i \Delta t$ represents the probability the node $i$ is active, where $a_i$ is the activity rate of node $i$. Without loss of generality we will assume $\Delta t=1$.
To get the ME for the in-degree $k_c$ distribution, we exploit also the time-dependence for a couple of passages:%\footnote{Also in this case is assumed $m=1$, for a generic value of m is enough to multiply each term in \emph{r.h.s.} per m.} 
\begin{widetext}
\begin{equation}
\begin{aligned}
P_c(s, k_c, t+ 1) &= P_c(s, k_c, t) \left[ 1 - \mu \sum_{j} a_j \right] + P_c(k_c, t) \left[  \frac{k_c}{s} \mu a_i + \frac{s-1}{s} \mu \sum_{j \nsim i} a_j + \mu \sum_{j\sim i}  a_j \right] + \\ 
& + P_c(s, k_c-1, t) \left[ \frac{s- k_c}{s} \mu a_i +\frac{1}{s} \mu \sum_{j \nsim i} a_j \right],
\end{aligned}
\end{equation}
\end{widetext}
where $\sum_{j \sim i}$ and $\sum_{j \nsim i}$ are respectively two contracted notations for the sum of all the first neighbors of node $i$ and the sum of all nodes but the neighbors of node $i$. The first parenthesis indicates the probability that none of the nodes in the network fire. Third (sixth) term is the probability a node $i$, in the instantaneous network, is active and fires to a node where, in the integrated counterpart, there is already (isn't) a link. Four (seventh) term is the probability a node $j$ not linked to $i$ fires to any of the nodes but $i$ (fires to i). Fifth factor is the probability a node $j$ already linked to $i$ fires.\\
After some algebra, ME can be written as:
\begin{widetext}
\begin{equation}
\begin{aligned}
P_c(s, k_c, t+1) - P_c(s, k_c, t) &= - \left[P_c(s, k_c, t) - P_c(s, k_c-1, t)\right] \Bigg(\frac{s-k_c}{s} \mu a_i + \frac{\mu}{s} \sum_{j\nsim i} a_j \Bigg).
\end{aligned}
\end{equation}
\end{widetext}
Now we pass to the continuum limit by considering $t\gg 1$ and $k \gg 1$. So the \emph{l.h.s} becomes simply the time derivative with respect to $P_c(s,k_c)$ and, to obtain a proper convergence of the results, we can expand the probability with respect to the incommunity degree up to second order. \\
In the regime $t\ll \tau(s)$, we can neglect $k_c \ll s$ and $ 1/s \sum_{j\nsim i} a_j \approx \left\langle a \right\rangle$. 
\begin{equation}
\frac{\partial P_c(s, k_c)}{\partial t} = \left( \mu a + \mu\left\langle a \right\rangle \right) \left[ \frac{\partial P_c(s, k_c)}{\partial k_c} - \frac{1}{2} \frac{\partial^2 P_c(s, k_c)}{\partial k_c^2} \right], 
\label{eq:ME_suppI}
\end{equation}
where we dropped the $a_i$ index since we expect all the nodes of a given activity to behave in the same way. Now $a$ is an activation rate and in the treatment we assume it takes small values to avoid that two nodes become active together.\\
The solution of Eq.~\ref{eq:ME_suppI} reads:
\begin{equation}
P_c(s, k_c) = C \exp\left[-\frac{(k_c - \mu(a+\left\langle a \right\rangle)t)^2}{2 \mu (a+\left\langle a \right\rangle)t}\right],
\label{P_c_final_equation_supp}
\end{equation}
where C is a normalization constant.\\
By following the same procedure, we recover the same results of Eq. \ref{P_c_final_equation_supp} for the out-community degree $k_o(a,t)$, by substituting $\mu\to\mu^{\prime}$ and $k_c\to k_o$:
\begin{equation}
P_o(s, k_o) = C \exp\left[-\frac{(k_o - \mu^{\prime}(a+\left\langle a \right\rangle)t)^2}{2 \mu^{\prime} (a+\left\langle a \right\rangle)t}\right]
\label{P_o_supp}
\end{equation}
Since $N \gg s_{max}$, the out-degree $k_o \ll N$ for any time $t$ of the process, thus we assume that Eq. \ref{P_o_supp} is valid for all the time scales analyzed. Also note that, as expected, the net effect of the mixing parameter $\mu^{\prime}$ is just a time rescaling of the out-community and in-community activity, respectively.\\
Then, in the $t \sim \tau_c(s)$ time range is not possible to find an analytic formula, however simulations will be run to provide, at least, a qualitative behavior. In the $t \gg \tau_c(s)$ time limit the $P_c(s, k_c)$ converges to the
$\delta(k_c - (s - 1))$ distribution. In fact, all the nodes will have all their edges activated and the $P_c(s, k_c)$ time derivative goes to zero. 

Let us now resume the results found in this section:
\begin{widetext}
\begin{subnumcases}{P_c(s, k_c) \propto}
    \exp\left[-\frac{(k_c - \mu(a+\left\langle a \right\rangle)t)^2}{2 \mu (a+\left\langle a \right\rangle)t}\right]  & for  $t \ll \tau_c(s)$ \label{P_c-small_supp}
    \\
    \delta(k_c - (s - 1))   & for $ t\gg \tau_c(s)$ \label{P_c-big_supp}
\end{subnumcases}
\end{widetext}

\begin{equation}
P_o(s, k_o) \propto \exp\left[-\frac{(k_o - \mu^{\prime}(a+\left\langle a \right\rangle)t)^2}{2 \mu^{\prime} (a+\left\langle a \right\rangle)t}\right] \quad \text{for } \forall t
\label{P_o_all_supp}
\end{equation}
where we now distinguish between the in-community degree distribution $P_c(s, k_c)$ and the out-community degree distribution $P_o(s, k_o)$. The latter however, is independent on the community size and we can then define $P_o(s, k_o) = P_o(k_o)$.\\
So far we treated the two probability functions separately when, in fact, $k_c$ and $k_o$ are bound by the relation
$k = k_c + k_o$. The total degree distribution $P(s, k)$ will then be determined by the convolution of both the
$P_c(s, k_c)$ and $P_o(s, k - k_c)$:
\begin{equation}
P(s, k) = \int_0^k dk_c P_c(s, k_c) P_o(k-k_c)
\label{convolution_supp}
\end{equation}
where we integrate over all the possible arrangements of the $k_c$ edges.\\
In the $t \ll \tau(s)$ limit, by substituting Eq. \ref{P_c-small_supp} and \ref{P_o_all_supp} in Eq. \ref{convolution_supp}, we sum the two exponents getting:
\begin{widetext}
\begin{equation}
P(s, k) = C \int_0^k dk_c \exp\left[-\frac{(k_c - \mu(a+\left\langle a \right\rangle)t)^2}{2 \mu (a+\left\langle a \right\rangle)t} - \frac{(k-k_c - \mu^{\prime}(a+\left\langle a \right\rangle)t)^2}{2  \mu^{\prime} (a+\left\langle a \right\rangle)t}\right]
\end{equation}

where C is, again, a normalization constant.\\
By combining the two terms and after some algebra we get:
\begin{equation}
P(s, k) = C \int_0^k dk_c \exp\left[-\frac{(k_c - \mu k)^2}{2 \mu \mu^{\prime}(a+\left\langle a \right\rangle)t} - \frac{k^2-2k(a+\left\langle a \right\rangle)t - (a+\left\langle a \right\rangle)^2 t^2}{2 (a+\left\langle a \right\rangle)t}\right]. 
\end{equation}

The integration over $k_c$ gives:

\begin{equation}
P(s, k) = C \left[ \text{Erf} \left(\frac{\mu^{\prime} k}{\sqrt{2 \mu \mu^{\prime} (a+\left\langle a \right\rangle)t}}\right)- \text{Erf} \left( \frac{\mu k}{\sqrt{2 \mu \mu^{\prime} (a+\av{a})t}}\right)\right] \exp\left[ - \frac{k^2-2k(a+\av{a})t - (a+ \av{a})^2 t^2}{2 (a+ \av{a})t}\right]
\label{Small_times_supp}
\end{equation}
\end{widetext}
where Erf$(x)$ is the error function evaluated at $x$.\\
In the small time limit, if we want to evaluate the $P(k) = \int_{s_{\text{min}}}^{s_{\text{max}}} ds P(s)P(s,k)$, we have to consider $t \ll \text{min}_s(\tau_c(s))$. In this way, for each community we can use Eq. \ref{Small_times_supp} as the true value of the $P(s, k)$. The integration over the different community size $s$ is then straightforward since the terms are independent on it, giving $P(k) = P(s, k)$. Note that this result holds for any value of $\mu$, $N$ and $a$.\\
The computation of $P(k)$ in the large time limit (i.e. $t \gg \left\langle \tau_c(s) \right\rangle$) is more complicated and we have to assume that $k_c = s-1$ for each node in a community of size $s$, otherwise Eq. \ref{P_c-big_supp} put everything equal to zero. The $P_o(s, k_o)$ will still be approximated by Eq. \ref{P_o_all_supp}. The integral now reads:
\begin{widetext}
\begin{equation}
\begin{aligned}
P(k) & = \int_{s_{\text{min}}}^{s_{\text{max}}} ds P(s)P(s,k)= C \int_{s_{\text{min}}}^{s_{\text{max}}} ds P(s) \exp\left[-\frac{(k - (s-1) - \mu^{\prime}(a+\left\langle a \right\rangle)t)^2}{2 \mu^{\prime} (a+\left\langle a \right\rangle)t}\right],
\end{aligned}
\label{P_big_times_supp}
\end{equation}

The exponential can be written as: 
\begin{equation}
\exp\left[-\frac{(k - \mu^{\prime}(a+\left\langle a \right\rangle)t)^2 + (s-1)[(s-1) -2(k-\mu^{\prime}(a+\left\langle a \right\rangle)t)))]}{2 \mu^{\prime} (a+\left\langle a \right\rangle)t}\right],
\end{equation}
\end{widetext}
where the rise of new terms proportional to $s^2$ and $sk$ makes it difficult to perform the integral.

We can, however, give a solution for the simple case $P(s) = \delta (s-\bar{s})$, when all the communities have equal size. First of all, we have, for large times, $P_c(s, k_c) = \delta(k_c - (s-1)).$ Then:

\begin{subnumcases}{P(s, k) =}
     P_o(k - (s-1))  & for  $k\geq s-1$ 
    \\
0   & for $ k<s-1$. 
\end{subnumcases}
When $k<s-1$, for sure $k_c < s-1$ and the delta put everything equal to zero. In the other case $k_c = s-1$ with a certain probability $P_o(k-(s-1))$.
Finally, equation \ref{P_big_times_supp} can be written as: 
\begin{subnumcases}{P(k) =}
     P_o(k - (\bar{s}-1))  & for  $k\geq \bar{s}-1$ 
    \\
0   & for $ k<\bar{s}-1$ 
\end{subnumcases}

% subsection ME and the $P(a, k, t)$ (end)

\subsection{The average degree $\av{k(a,s,t)}$} % (fold)
\label{sub:average_degree$}
We can provide a simple expression for the nodes average degree belonging in different classes. As we already showed in Eq. \ref{k_c_supp}, $\left\langle k_c (a,s,t) \right\rangle$ grows as:
\begin{equation}
\begin{aligned}
\left\langle k_c(a,s,t) \right\rangle &= (s - 1) \left(1 - \exp\left(- \frac{t}{\tau(s)} \right)\right) = \\ &= (s-1)\left(1-P^{\prime}(s)\right)= \\
&=(s-1)\left[1-\left(1-\frac{c}{s-1}\right)^t \right],
\end{aligned}
\end{equation}
and the $\left\langle k_o \right\rangle$ grows as the mean value of the distribution given in equation \ref{P_o_all_supp}, and turns out to be independent on $s$.
\begin{equation}
\left\langle  k_o(a,t) \right\rangle= \mu^{\prime}(a+\left\langle a \right\rangle)t 
\end{equation}
The average total degree $k(a,s,t)$ for nodes of activity $a$ belonging to communities of size $s$ depends on the time scale we analyse the problem. For small times (i.e. $t \ll \left\langle \tau(s) \right\rangle_s$):
\begin{widetext}
\begin{equation}
\begin{aligned}
\left\langle k(a,s,t)\right\rangle &= \left\langle k_c(a,s,t)\right\rangle +\left\langle k_o(a,t) \right\rangle\approx\\
&\approx (s-1)\left(1 - 1 + \frac{\mu(a+\left\langle a \right\rangle)t}{s-1} \right) + (1-\mu)(a+\left\langle a \right\rangle)t = \\
&= (a+\left\langle a \right\rangle)t
\end{aligned}
\label{k_T short time_supp}
\end{equation}
\end{widetext}
As time grows toward the regime of times comparable to the average (i.e. $t \sim \left\langle \tau (a,s,t) \right\rangle_s$), we cannot approximate the in-community degree anymore but we use directly equation \ref{k_c_supp}:
\begin{equation}
\left\langle k(a,s,t)\right\rangle = (s - 1) (1 - P^{\prime}(s)) + \left\langle k_o(a,t) \right\rangle.
\label{k_similar_times_supp}
\end{equation}
Then, the regime of large times (i.e. $t \gg \left\langle \tau (s) \right\rangle_s$) is:
\begin{equation}
\left\langle k(a,s,t)\right\rangle \approx s-1 +  \mu^{\prime} (a+\left\langle a \right\rangle)t  \approx \mu^{\prime} (a+\left\langle a \right\rangle)t 
\end{equation}
The above equations then predict that the average degree has a linear growth proportional to $a+\left\langle a \right\rangle$ for short time limit (equation
\ref{k_T short time_supp}), then a transition for $t \sim \left\langle \tau (s) \right\rangle_s$ is followed by a second linear growth, valid for large times, proportional to $\mu^{\prime}(a+\left\langle a \right\rangle)$. These regimes correspond to the initial growth, in which both the in-community and the out-community degrees are growing linearly in time, followed by the slowing down of the in-community degree which is saturating to $s - 1$. Finally, the third regime is again linear and it is driven by the $\mu^{\prime}(a+\left\langle a \right\rangle)$ coefficient: it means meaning that only the out-community degree is growing.

% subsection The average degree $\mean$ (end)

\subsection{The degree distribution $\rho(k_{\text{th}})$} % (fold)
\label{sub:degree_distribution}
Now that we have the expression of the average degree, it is straightforward to write the degree distribution. At all the time scales we found $\left\langle k_{\text{th}} \right\rangle \propto Ct$, where $C$ is a time-independent coefficient. Then, $k_{\text{th}} \propto at$, and it results in an equal increment in the activity and in the degree values ($da = dk_{\text{th}}$). If we use the change of variable rule, we obtain:
\begin{equation}
F(a)da = \rho(k_{\text{th}}) dk_{\text{th}} \Rightarrow  a^{-\nu}da = k_{\text{th}}^{-x}dk_{\text{th}} \Rightarrow x = \nu,
\label{Theoretical_degree_distribution_supp}
\end{equation}
i.e., the degree distribution has the same exponent $\nu$ as the activity distribution function.

%For $t \gtrsim \av{\tau(\aai, s)}_s$ instead, we found $k(a, t) = (1 - \mu)(a +
%\av{a})t + \av{s}$.
%Then:
%\begin{eqnarray}
%    a =& \frac{k - \av{s} - (1 - \mu)\av{a}t}{(1 - \mu)t} \;\; \Rightarrow
%    \;\; da =& \frac{dk}{(1 - \mu)t} \nonumber \\
%    F(k) &= \left( \frac{k - \av{s} - (1 - \mu)\av{a}t}{(1 - \mu)t}
%    \right)^{-\nu} \frac{dk}{(1 - \mu)t}. &\;
%    \label{eq:rhok2}
%\end{eqnarray}

% subsection The degree distribution $\rho(k)$ (end)

% section The Network growth (end)

\section{Comparison with numerical simulations} % (fold)
\label{sec:simulations}

To check the analytical predictions of Section \ref{sec:Network_growth} we
performed numerical simulation. In particular we realized $100$ representations
of a network featuring:
\begin{itemize}
    \item [-] $N=10^5$ nodes with modularity $\mu=0.9$ evolving for $10^5$
        evolution steps;
    \item [-] activity potential distributed following the $F(a)\propto
        a^{-\nu}$ with $\nu=2.1$ and $a\in[10^{-3}, 1]$ interval;
    \item [-] power-law distributed community sizes $P(s)\propto s^{-\omega}$
        with $\omega=2.1$ and $s\in[10,\sqrt{N}]$.
\end{itemize}

In order to analyze the collective behavior of the nodes we group them by their
activity and community size, thus defining $b$ classes of nodes.
We average over the representations of the network and for each class of nodes
$b$ we evaluate:
\begin{itemize}
    \item [-] $P_c(s,k)$, $P_o(k)$, $P(s,k)$ for $t\ll \tau_c(s)$ and
        $t\gg \tau_c(s)$;
    \item [-] the average degree $\av{k_c(s)}$, $\av{k_o}$ and
        $\av{k(s)}$;
    \item [-] the degree distribution $\rho(k_{\text{th}})$.
\end{itemize}

\begin{figure}
\centering
\includegraphics[width=9.2cm]{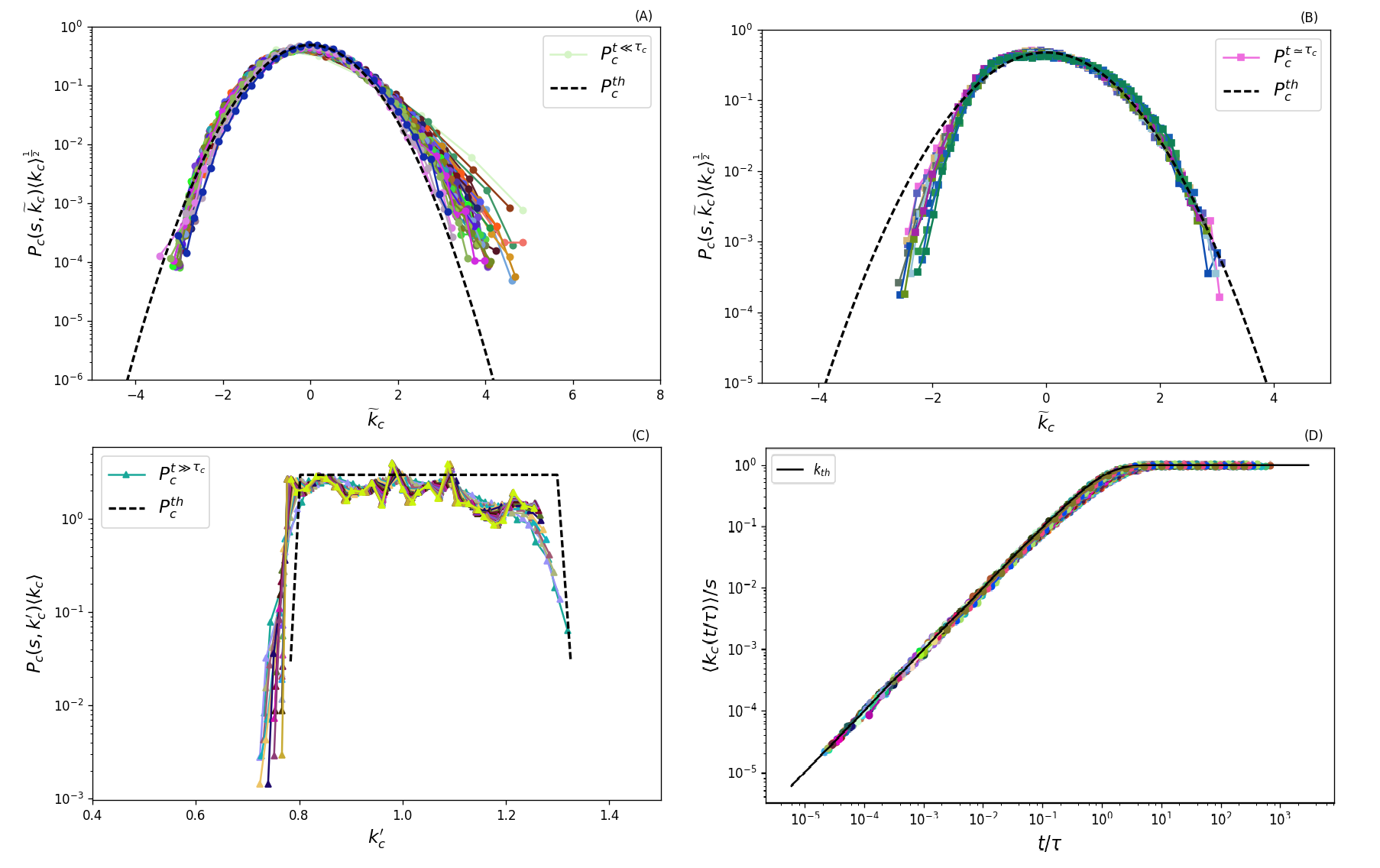}
\caption{Panels A-C) rescaled $P_c(k_c)$ probability distribution as found for a selected node class at different times (legends). Functions in panels A-B) are rescaled accordingly to the theoretical distribution given in Eq. \ref{P_c-small_supp}, i.e., by sending $k_c \rightarrow \tilde{k_c} = (k_c - \av{k_c})/\av{k_c}^{1/2}$ and plotting $P_c(\tilde{k_c}) \av{k_c}^{1/2}$. Panel C) is rescaled accordingly to the theoretical distribution given in Eq. \ref{P_c-big_supp}, i.e., by sending $k_c \rightarrow k_c^{\prime} = k_c/\av{k_c}$ and plotting $P_c(k_c^{\prime}) \av{k_c}$. In panel D) we plot $\av{k_c}$ for nodes featuring different activity potential and belonging to communities of different size. The data are rescaled sending the time $t \rightarrow = t/\tau$ and then plotting $\av{k_c(t/\tau)}/s$. The analytical prediction of Eq. \ref{k_c_supp} computed for a node of activity $a = \av{a}$ and belonging to a community of size $s = \av{s}$. Each point is an average over $10^2$ independent simulations, parameters used are: $N = 10^5$, $\omega = 2.1$, $\nu = 2.1$, $m=1$, $s_{min}=10$ and $\mu = 0.9$}
\label{P_c_k_c_plots_supp}
\end{figure}

In the main discussion, we already showed that some of the above measures have a great agreement with analytical predictions. To complete the discussion, we add below Fig. \ref{P_c_k_c_plots_supp} which proves that also the in-community probability and the in-community degree perfectly matches our expectation. In panel B) we display $P_c(s,k)$ for $t \sim \tau_c$ even if it was impossible to obtain an exact result, the comparison demonstrates that the in-community probability starts to deviate from a Gaussian distribution. 

\newpage
\section{SIR and SIS processes on modular activity driven networks}
We present together all the results about SIR and SIS models obtained in synthetic networks. We start our simulations by setting $1\%$ of randomly selected nodes as initial infected seeds, the other parameters are fixed as $\gamma=0.01$, $m = 3$, $\nu = 2.1$, $N=10^5$ and  $Y=0.5$. Panels A) and C) set the exponent of the distribution of community sizes $\omega=1.5$, while panels B) and D)  $\omega=2.1$ (they are respectively Fig.\ref{SIR}B and \ref{SIS}B). The qualitative picture is unchanged due to selecting a different value of $\omega$. The modular structure becomes irrelevant for $\mu \rightarrow 0$, whilst it significantly modifies the spread of the disease when $\mu \gg 0$. Modularity slows down contagion processes in SIR models, it favors the disease outbreak in SIS models. Moreover, quantitatively, in SIR models the presence of larger communities lead to higher epidemic size, in SIS models it lower the epidemic threshold. Moreover, in SIS models and when $\mu \rightarrow 1$, the epidemic threshold is likely to increase due to major limitations in reaching an endemic size. This phenomenon is particularly visible in Panel C), solid blue curve.    

\begin{figure}[h]
\centering
\includegraphics[width=9cm]{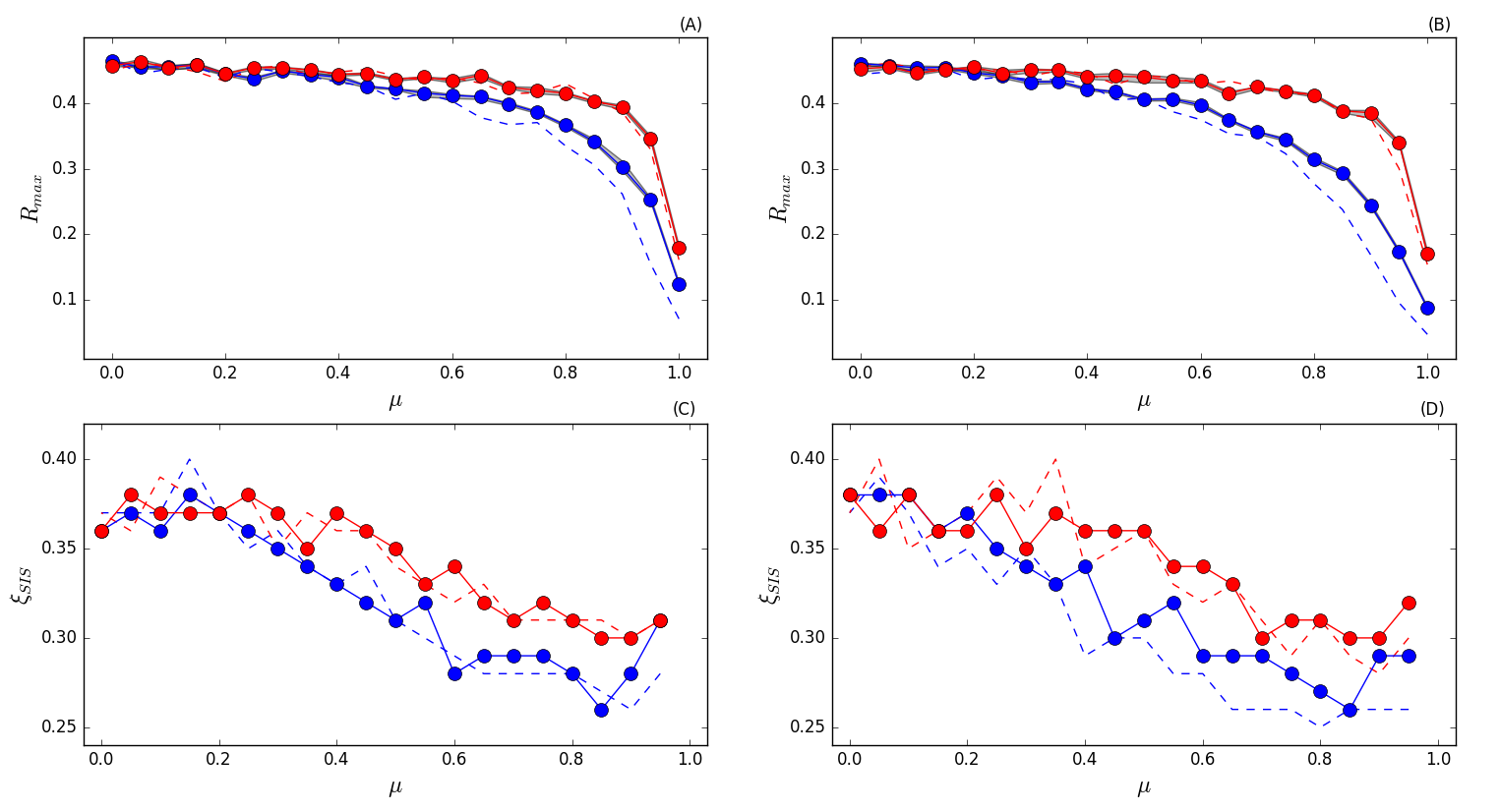}
\caption{Panels A-B) take $R_{max}$ of each $R_{\infty}$ curve and plot it as a function of $\mu$.  Panels C-D) $\xi_{SIS}$, that is $\beta/\gamma$ in correspondence of $L_{max}$, and plot it as a function of the modularity. In red curves we set $s_{min}=100$, in blue curves $s_{min}=10$. Solid curves are obtained by drawing community sizes from a power law distribution, 95\% confidence interval is in gray. Dashed curves have a constant community size equal to the average value of the power law. Each point is an average of $10^2$ independent simulations.}
\label{SIS_SIR}
\end{figure}

\newpage
\section{Real network: APS dataset} 
\begin{figure}[h]
\centering
\includegraphics[width=6cm]{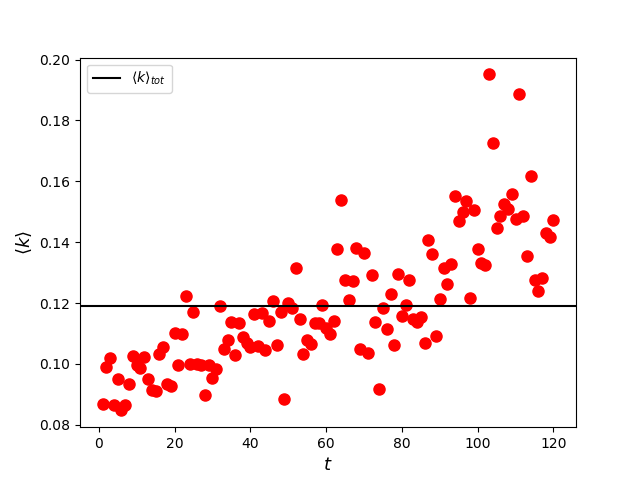}
\caption{Average degree for each month in the selected subset of APS dataset (red circles), compared with the global average of all the ten years considered (solid black line). We label each month with increasing integer numbers from 1 to 120, where 1 represents the beginning of our sample, January 1997, and 120 the end, December 2006. We observe an increasing number of collaborations through years.}
\label{av_k_supp}
\end{figure}

In the data each author of an article is described as a node. An undirected link between two different authors is drawn if they collaborated in the same article. We used the dataset from Ref \cite{SARA} which spans a period between $1893$ and $2006$. To have the average degree in each instantaneous network as comparable as possible (see Fig.  \ref{av_k_supp}), we select a period of ten years, from January 1997 to December 2006. In this time window we register $96940$ scholars who create $692667$ connections. When we simulate SIR and SIS models on top of APS temporal network, we use periodic boundary conditions to let the disease dynamics evolve without late-time constrains. \\

\begin{figure}[h]
\centering
\includegraphics[width=9cm]{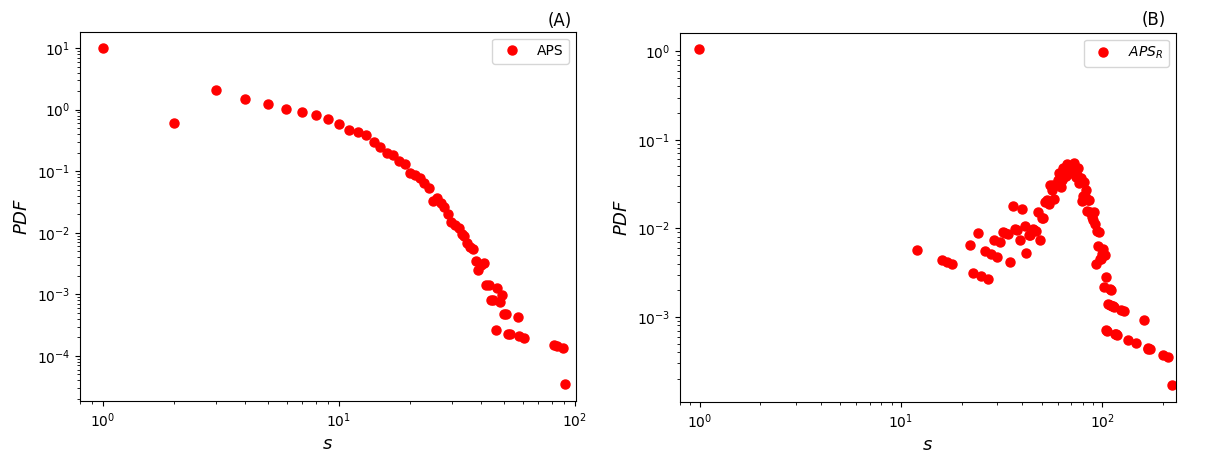}
\caption{Panel A) Community sizes probability density function in the original network. Number of communities found is $10825$, minimum community size is $1$, maximum $91$. Panel B) Community sizes probability density function after having applied the degree-preserving randomization. Number of communities found is $1489$, minimum community size is $1$, maximum $222$. Note also that the shape of the distribution is completely different from panel A).}
\label{pdf_size}
\end{figure}

Using OSLOM, in Fig.\ref{pdf_size}A we show that the integrated APS network is modular. Then, we apply the following degree-preserving randomization technique to destroy the network's community structure. We choose randomly a source node $S_1$ and, among its neighbors, we select randomly a target node $T_1$. We do the same for other two nodes $S_2$ and $T_2$. If the two pairs are equivalent or if a multi-edge will be created, we start back by selecting $S_1$ again and repeating the instructions. Otherwise, we swap $T_1$ and $T_2$ to have the new undirected links: $S_1-T_2$ and $S_2-T_1$. The edges are chosen within the same temporal network and the number of swaps is equal to the number of edges in that instantaneous network. So, the above procedure is applied for each instantaneous network. At the end, we integrate the randomized temporal network and use OSLOM to detect the community structure. In Fig.\ref{pdf_size}B, we qualitatively prove that our degree preserving randomization destroys the network's community structure.\\
The options used in OSLOM are: -uw (to study undirected networks); -cp0.99 (to have communities as large as possible); -hr (to avoid to consider hierarchies); -r100 (to repeat 100 times the community detection). Since OSLOM finds communities in a non-deterministic way, last option is useful to get rid of stochastic fluctuations and have a more reliable community structure.\\
Finally, we evaluate quantitatively the modularity of the two networks. For the original APS network, $Q=0.6685$, and for its randomized counterpart $Q =0.0937$.

\end{document}